\documentstyle[psfig]{article}
\renewcommand{\Re}{{\rm Re}\,}
\begin{document}
\begin{flushright}
 preprint THU-93/22\\[10mm]
\end{flushright}
\begin{centering}
 {\bf\LARGE Instabilities and Patterns}\\[10mm]
 M.H. Ernst and H.J. Bussemaker\\
 Institute for Theoretical Physics\\
 University of Utrecht, The Netherlands\\[10mm]
\end{centering}
\begin{abstract}
Violation of (semi)-detailed balance conditions in lattice gas automata
gives rise  to unstable spatial fluctuations that lead to phase
separation and pattern formation in spinodal decomposition, unstable
propagating modes, driven diffusive systems and unstable uniform flows.\\[5mm]
\end{abstract}

\section{Introduction}

Why is violation of detailed or semi-detailed balance \cite{Stueck} of
interest, and what are its consequences? In the context of lattice gas
automata violation of the Stueckelberg condition has been used to increase
the Reynolds number in hydrodynamic flows \cite{Hen-Nice} and to model the
dynamics of phase separation and spatial instabilities
\cite{RK-color,AZ,Gerits,Alex}. Lack of detailed balance occurs in several
areas of statistical and chemical physics, such as driven diffusive systems
\cite{Zia}, in Smoluchowski's theory of rapid coagulation
\cite{Drake,ernst} and in the theory of neural networks and pattern
recognition \cite{Coolen,Derrida}, where master equations with asymmetric
transitions rates are being used.

As soon as detailed balance is lacking, there does not exist an
H-theorem. In the thermodynamic limit the system  may or may not
approach a unique stable spatially uniform equilibrium state. In the
latter case the system seems to build up spatial correlations of 
ever-increasing range \cite{Berlin}; in the former case there exist strong
velocity correlations  at equal times \cite{Hen-Nice,Nice}. Very
little is known about the possible stationary states and further
properties of non-detailed balance models.

In this paper we study lattice gases that violate the Stueckelberg
condition, using mean field theory. We are particularly interested in
spatial instabilities, that lead to phase separation and formation of
spatial patterns. The main goal is to show how a linear stability
analysis or study of the long wavelength hydrodynamic and diffusive
modes determines the stability thresholds (temperature, range of
forces), the initial patterns, i.e. wavelength and direction of maximum
growth, and the onset time of spatial instabilities. The coarsening of
the initial patterns, the growth of correlation length and the scaling 
laws describing the late stages of phase separation can not be explained 
on the basis of this linear theory.

The paper is organized as follows. In section 2 we review the lattice
Boltzmann equation, the decay or growth of respectively stable or
unstable spatial fluctuations, and the Cahn-Hilliard theory. Section 3
presents various mechanisms for phase separation in lattice gas
automata: thermodynamic instabilities, unstable propagating modes and
fluctuating bias fields. In section 4 the discussion is extended to
instabilities in systems where the spatial symmetry is externally broken
by an applied field, i.e.\ driven diffusive systems, or by an imposed
uniform flow velocity.

\section{Linear stability analysis}

\subsection{Lattice Boltzmann equation}

Starting from the nonlinear Boltzmann equation  for lattice gas automata
with local and nonlocal collision rules, we investigate the linear
stability of long wavelength fluctuations around the spatially uniform
state.
If the stability criterion is not satisfied, long wavelength
fluctuations 
(e.g.\ concentration fluctuations in spinodal decomposition) are unstable
and the Cahn-Hilliard theory is used to discuss the initial behavior of 
the structure function and the density-density correlation function.
In standard notation the lattice Boltzmann equation is given by,
\begin{equation}\label{a1}
  f_i(\vec r + \vec c_i,t +1) -  f_i(\vec r,t) = I_i(f(t))
\end{equation}
where  $i=1,2,\ldots,b$ denotes a moving particle state with $|\vec c_i|
=v_0$  and $i=0$ denotes a rest particle state with velocity $\vec c_0
=0$. 
The collision operator $I_i$ contains in general not only the standard
local collisions, but also nonlocal ones.

We are interested in the stability of long wavelength spatial
fluctuations of the form  \cite{Berlin} ,
\begin{equation}\label{a2}
  \delta f_i(\vec r,t) = f_i(\vec r,t) -f^o_i = \exp[i\vec k \cdot \vec r
                         + z_\lambda(\vec k)t]
\end{equation}
around the spatially uniform state $f^o_i$, given by the solution of
$I_i(f^o)=0$. Inserting (\ref{a2}) into (\ref{a1}) yields after
linearization an eigenvalue equation for the $\lambda$-th mode $\tilde
\psi_{\lambda j}$ with eigenvalue $z_{\lambda}$, i.e.
\begin{equation}\label{a3}
  \exp[z_\lambda]  \tilde \psi_{\lambda i}(\vec k) = \sum_{j=0}^b
  \exp[-i\vec k \cdot \vec c_i](1+\Omega(\vec k))_{ij} \tilde
  \psi_{\lambda j}(\vec k)
\end{equation}
where $\Omega(\vec k) \simeq \Omega + ik \Lambda \ldots (\vec k \to 0)$
is the Fourier transform of the linearized nonlocal collision operator
$I_i(f^o+\delta f)$.

What type of information can be obtained from Eq.(\ref{a3})? We mainly
focus on the eigenvalue spectrum $z_\lambda(\vec k)$ and in particular
on the soft hydrodynamic modes where  $z_\lambda(\vec k) \to 0$ as 
$\vec k \to 0$, that are related to the conservation laws. If $\Re
z_\lambda(\vec k) >0$ the mode is unstable and the corresponding
eigenfunction $\tilde \psi_\lambda$ is the order parameter.

The eigenvalues $z_\lambda(\vec k)$ can be determined either numerically
or analytically by perturbation theory for small $|\vec k| = k$. In the
long wavelength limit the hydrodynamic eigenvalues have the form,
\begin{equation}\label{a4}
  z_\lambda(\vec k)=ik v_\lambda + (ik)^2 \Gamma_\lambda + \ldots
 \qquad (\vec k \to 0)
\end{equation}
In a single component $d$-dimensional lattice gas with mass and
momentum conservation, there are ($d+1$) slow modes: two {\em
propagating} sound modes (labeled $\lambda = \sigma= \pm$), where
$\Gamma_\sigma = \Gamma$ is the sound damping constant, and $v_\sigma =
v_s$ is the speed of sound, defined through $v^2_s = dp/d\rho$. The
remaining $(d-1)$ shear modes (labeled $\lambda = \perp$) are degenerate
where $\Gamma_\perp = \nu$ is the kinematic viscosity and $v_\perp =0$.
In a binary mixture $\Gamma_D = D$ is the diffusion coefficient.

\subsection{Hydrodynamic modes}

We first consider the analytic method. By applying the perturbation
theory of Gerits et al.\ [1993] to nonlocal collision operators, $\Omega(\vec k)
=  \Omega + ik \Lambda + \ldots $, the damping coefficient of the
$\lambda$-th mode is found to have the form,
\begin{equation}\label{a5}
  \Gamma_\lambda = \left( \frac{1}{\omega_\lambda}- \frac{1}{2} \right)
  \langle j_\lambda | \tilde j_\lambda \rangle - \frac{1}{\omega_\lambda}
  \langle j_\lambda | \Lambda | \tilde a_\lambda \rangle
\end{equation}
or a linear combination of such terms. We have used the notation
$\langle a|b \rangle = \sum_i a_i b_i$ and 
$\langle a|M|b \rangle = \sum_{ij} a_i M_{ij} b_j$.
The current $\tilde j_\lambda$ is a right eigenvector of the collision
matrix, $\Omega \tilde j_\lambda = -\omega \tilde j_\lambda$ with 
$0 < \omega_\lambda < 2$, and $j_\lambda$ is the left eigenvector
corresponding to the same eigenvalue; $\tilde a_\lambda$ and $
a_\lambda$ are the zero-eigenfunctions (collisional invariants) with
$\omega_\lambda=0$, normalized as 
$\langle a_\lambda|\tilde a_\lambda \rangle =1$.
Here left and right eigenvectors are different because the matrix
$\Omega $ is asymmetric in lattice gas automata that violate the
condition of detailed balance. The vectors $a_\lambda$ and $j_\lambda$
are related as $j_\lambda = Q c_\ell a_\lambda$, where $c_\ell = \hat k
\cdot \vec c$ and $Q$ projects out the components of $c_\ell a_\lambda$
parallel to the collisional invariants. In case the collision rules are
purely local the matrix $\Lambda$ vanishes. The sound damping constant
$\Gamma$ is a linear combination of kinematic viscosity $\nu$ and bulk 
viscosity $\zeta$ given by,
\begin{eqnarray}\label{a6}
  \Gamma & =& (1-{1\over d})\, \nu + \textstyle{\frac{1}{2}} \zeta
  \nonumber\\
  \nu & = & \left( \frac{v_o^2}{d+2} \right) \left(
  \frac{1}{\omega_\nu} -  \frac{1}{2}\right) \nonumber \\
  \zeta & =& \left( \frac{v_o^2}{d}- v_s^2 \right) \left(
  \frac{1}{\omega_\zeta}-\frac{1}{2}\right).
\end{eqnarray}
In the standard detailed balance models the eigenvalues $\omega_\lambda$
are simple functions of the density. In models violating the
Stueckelberg conditions,  such as the FCHC lattice gas \cite{Coev} or
the biased triangular lattice gas \cite{Berlin}, the eigenvalues have to 
be calculated numerically. Although no H-theorem can be proven for these
models, it turns out that the eigenvalues still satisfy the
inequalities, $0 < \omega_\lambda <  2$, although transport
coefficients may be negative (see section 3.2).

\begin{figure}[t]
\centerline{ 
 \psfig{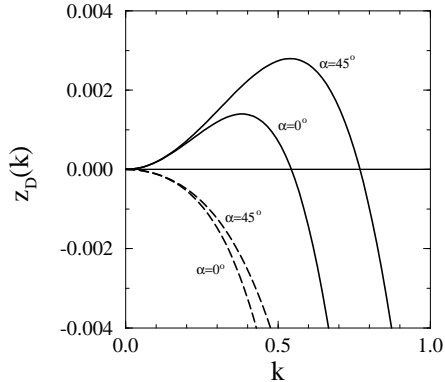}
}
\caption{Damping constant $z_D(\vec k)$ of diffusion mode versus 
 $\vec k$ in negative diffusion model \protect{\cite{Alex}},
 where $\alpha$ is the angle between $\vec k$ and $\vec c_1$, the $\hat
 x$-axis. Density $\rho=2.0$ and threshold $T_c \simeq 3.8$. At $T=4.0$
 (dashed lines) the modes are stable; at $T=3.5$ (solid
 lines) the modes are unstable.
}
\end{figure}

Next we turn to the numerical solution of the eigenvalue equation
(\ref{a3}). For a given set of collision rules we evaluate the matrix
elements $\Omega_{ij}(\vec k)$ and the eigenvalues $z_\lambda(\vec k)$
numerically. Typical results are illustrated in Fig.~1 for a square
lattice gas, described in section 3.3,  with a single slow (diffusion)
mode. The figure shows the damping constant  $z_D(\vec k)$ of the
diffusion mode as a function of $\vec k$. The diffusion mode is only
{\em  stable} (dashed lines) if the control parameter $T$ (defined in
section 3.3) exceeds a certain threshold value $T_c$. For $T < T_c$
(solid lines in Fig.~1) the modes are unstable in two independent $\vec
k$-directions.

\subsection{Cahn-Hilliard theory}

What is the important information that can be obtained from the spectrum
$z_\lambda(\vec k)$ regarding stability of fluctuations and behavior of 
structure function and density-density correlation function?
The {\em criterion} for dynamic stability of long wavelength
fluctuations is $ \Re z_\lambda(\vec k) < 0$, which reduces to the
criterion $ \Gamma_\lambda > 0$ according to (\ref{a4}) in cases where
$v_\lambda$ is real. Furthermore, if there is a control parameter, say
$T$, then the root $T_c$ of the equation, $ \Re z_\lambda(\vec k,T) =0$,
determines the  threshold value $T_c$, below/above which the
$\lambda$-th mode is unstable/stable.

Numerical evaluation of the unstable eigenvalue $z_\lambda(\vec k)$
provides the following additional information:
(i) the $\vec k$-direction of maximum instability (in Fig.~1: $\alpha=45^o$);
(ii) the cut-off wavelength $\lambda_c = 2 \pi/k_c$ above which all
modes are unstable (in Fig.~1: $k_c \simeq 0.77$ at $\alpha=45^o$);
(iii) the wavelength of maximum growth $\lambda_m =2 \pi/k_m$ (in Fig.~1:
$k_m \simeq 0.55$ at $\alpha=45^o$);
(iv) the onset  time of instability $\tau_o = [\Re  z_\lambda(\vec
k_m)]^{-1}$ (in Fig.~1: $\tau_o \simeq 350$ at $\alpha=45^o$). 
The corresponding eigenfunctions are the order parameters, i.e.\
in shear modes it is the
transverse momentum density and in sound modes it is a linear combination
of longitudinal momentum and mass density.

The existence of unstable long wavelength modes indicates that the
system starts to phase separate and develop spatial structure on a
typical initial length scale $\lambda_m$. The $\vec k$-directions of
maximum growth determine the patterns in the structure function. Once
eigenvalues and eigenfunctions have been computed, and possible unstable
modes with $ \Re  z_\lambda(\vec k) >0$ have been identified, one can
use the Cahn-Hilliard theory to evaluate the structure function
$S(\vec k,t)$ or its inverse Fourier transform, the density-density
correlation function $G(\vec r,t)$,
\begin{equation}\label{a7}
  S(\vec k,t) = V^{-1} \langle |\rho(\vec k,t)|^2 \rangle 
              = \sum_{\vec r} \exp[-i\vec k \cdot \vec r\,] G(\vec r,t)
\end{equation}
where $V$ is the number  of nodes in the system, and $\rho(\vec k,t)$
the Fourier component of the microscopic density fluctuation  $\delta
\rho(\vec r,t) = \rho(\vec r,t) - \rho$ around the uniform density 
$ \rho = \sum_i f^o_i$. 
Its long wavelength components develop essentially like the
hydrodynamic modes $\rho(\vec k,t) \sim \exp [ z_\lambda
(\vec k) t]$. Hence the structure function at the onset of instability
is determined by the unstable long wavelength modes, i.e.
\begin{equation}\label{a8}
  S(\vec k,t) \simeq \left\{ \begin{array}{ll}
  S(\vec k,0) \exp[2\Re z_\lambda(\vec k)t] \qquad & (k < k_c)  \\
          S(\vec k,0) &(k > k_c)
  \end{array}
  \right.
\end{equation}
The arguments presented here are essentially the Cahn-Hilliard
theory of spinodal decomposition \cite{Langer}. We assume that
the structure function $S(\vec k,t)$ $=S(\vec k,0)$ remains in
equilibrium for $k>k_c$.

\section{Phase separation}

\subsection{Thermodynamic instabilities}

We consider a lattice gas  model for a gas-liquid system with an
attraction of (long) range $R$, which is the control parameter. It is
essentially a 7-bit FHP model to which a square well attraction of range
$R$ has been added. If the Fermi exclusion principle permits, there is an
instantaneous exchange  of momentum between two particles a distance $R$
apart, such that their relative velocity is always reduced. The model
therefore violates detailed balance. The momentum flux contains not only
a kinetic part, but also a collisional transfer part.

The perturbation theory of section 2 has been applied to this gas-liquid
model \cite{Gerits}. The pressure $p$, the speed of sound $v_s$ and the 
two viscosities $\nu$ and $\zeta$ have been calculated analytically, and 
have been succesfully compared with computer simulations. 
For the present discussion only the equation of state,
\begin{equation}\label{b1}
  p=3f - 2R f^2 (1-f)^2,
\end{equation}
is of interest, as it shows the thermodynamic instability. Here $6f=\rho
- f_0$ is the average number of moving particles per node, and $f_0$
that of rest particles. For $R>R_c=8$ the equation of state exhibits a
van der Waals loop. In the density interval (spinodal region) where
$dp/d\rho = v^2_s < 0$, the spatially uniform state is
thermodynamically  unstable. In computer simulations one observes
droplet formation. The measured pressure $p$ at short times ($t < 500$)
is in agreement  with (\ref{b1}) for all densities. For $t>500$ one
observes the coexistence pressure $p_c$. 

Inside the spinodal region ($1.7 < \rho < 2.6$) the pressure $p_c$
is in excellent agreement with the pressure obtained from Maxwell's
equal area construction. In the metastable regions, where nucleation is
the basic mechanism for droplet formation, the coexistence pressure
$p_c$ could only be observed after artificially creating some high or
low density droplets, acting as nuclei of condensation. 

The spinodal decomposition can also be analyzed in terms of the dynamic
instabilities of section 2. Here the speed of sound is purely imaginary,
$\Re z_\sigma(\vec k) = \pm |a|k$. Hence the sound modes stop to be
propagating, and the amplitude of one of them starts to grow as $\sim
\exp[ |a|kt]$. Also in a continuous gas-liquid system spinodal
decomposition is closely linked  to the instability of sound modes
\cite{Desai}.

\subsection{Unstable propagating modes}

The {\em biased triangular} lattice gas, introduced by Bussemaker 
and Ernst [1993a],
shows a dynamic instability with unstable sound modes that remain
propagating. The model is a stochastic 7-bit lattice gas  with {\em
strictly local}\, collision rules, that preserve the lattice symmetries,
but violate detailed balance. Typically, the forward and backward
collision probabilities are different whenever a rest particle is
involved.

\begin{figure}[t]
\centerline{
 \psfig{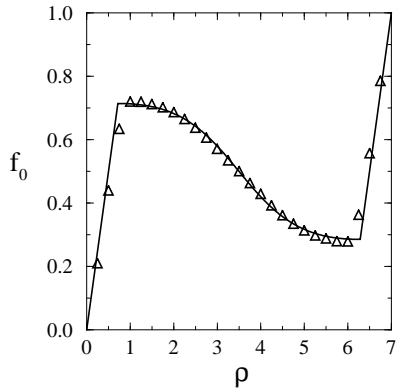}
}
\caption{Density of rest particles $f_0( \rho)$ versus $\rho$ 
 in the biased triangular model. Mean field theory (solid line) and 
 simulations (triangles) \protect{\cite{Berlin}}.
}
\end{figure}

The most characteristic feature of this model is shown in Fig.~2. It has
been observed in computer simulations, and confirmed by numerical
solution of the nonlinear lattice Boltzmann equation (\ref{a1}). There
is a density interval, where the the spatially averaged density of rest
particles $f_0(\rho)$ has a {\em negative slope}. In this model the
pressure is purely kinetic,
\begin{equation}\label{b2}
  p=3f= \textstyle{1\over 2}[ \rho - f_0(\rho)\,]
\end{equation}
and the speed of sound $v_s$, given by $v_s^2 = {1\over 2} [  1-
df_0(\rho)/d \rho]$, can exceed the value $1/\sqrt{2} \simeq 0.71$. In
the half-filled lattice $v_s \simeq 0.76$, as can be read off from
Fig.~2.
Consequently the bulk viscosity in the biased triangular lattice gas 
is {\em negative}\, on account of (\ref{a6}).

A negative bulk viscosity by itself is not a sufficient condition for
instability, as can be concluded from the work of Bussemaker and 
Ernst [1992], where several stable
lattice gases with the characteristics of Fig.~2 have been simulated. The
sound modes only become unstable if the sound damping constant $ \Gamma
={1 \over 2} (\nu + \zeta)$ becomes negative, i.e. if in addition $\nu$
is sufficiently small. These conditions are realized  in the biased
triangular lattice gas.

More quantitative information is obtained  by  solving  the eigenvalue
problem. One finds a density interval where sound modes are unstable.
For instance, at  $\rho=3.5$ one has $ \Re z_\sigma (\vec k) >0$ for $k
<  k_c$ with $k_c \simeq 0.6$; it reaches a maximum value $\tau_o^{-1}
= \Re z_\sigma (k_m) \simeq 0.005$ at $k_m \simeq 0.4$ \cite{Berlin}.

The onset of phase separation, the coarsening phenomena, the late stages
of growth, the scaling  properties of the structure function $S(\vec k,t)$ 
for the different order parameters, and the exponent $\alpha$ of
the dynamic correlation length $\lambda_m \sim t^\alpha$ in this model, 
have been extensively discussed in the literature \cite{Berlin,PhysLett}.

\subsection{Fluctuating bias fields}

The best known model of this kind is the Rothman-Keller model
\cite{RK-color} of a binary fluid consisting of red and yellow particles
({$\sigma =r,y$}), which are mechanically identical. There is an
attraction between like particles to enhance phase separation. The basic
model is again a triangular lattice gas with seven different velocity
channels and at most one particle per channel, i.e. the occupation  
number $s_i(\vec r)=0 $ or 1. To indicate the color $(\sigma =r,y)$ one
adds a color label to the occupation number $s_{\sigma i}(\vec r)$ with
$s_{r i}(\vec r) + s_{y i}(\vec r) = s_{ i}(\vec r) $.

The transition probability $A(s \to s^\prime)$ from a pre-collision
configuration $s(\vec r) = \{ s_{\sigma i}(\vec r); \sigma =r,y;
i=0,1,2,\ldots,b\}$ at node  $\vec r$ to a post-collision configuration
$s^\prime (\vec r)$ contains the usual delta functions accounting for
the conservation per node of red and yellow particles and of total
momentum. In addition there is a bias factor $\exp[\, \beta \vec
J(s^\prime) \cdot \vec G( {\bf s}_{n.n.})\,]$. The normalization is fixed by
the condition $\sum_{s^\prime} A(s \to s^\prime) = 1$. The bias factor
represents the effect of a local  bias field $ \vec G( {\bf s}_{n.n.})$,
which is here a color gradient, that depends on the configurations $\{
{\bf s}_{n.n.} \}$ of all nearest neighbor nodes. It further contains  the
color flux $ \vec J(s^\prime)$ in the post-collision configuration $
s^\prime (\vec r)$.

The quantity $\Delta W = \vec J(s^\prime) \cdot \vec G( {\bf s}_{n.n.})  $
represents the amount of work done on the system by sending particles up
a color gradient. The control parameter $T =1/\beta$ is a {\em
temperature-like} variable. The relative probability for  the transition
$s \to s^\prime$ is given by the bias factor $\exp[\beta \Delta W]$.
Depending on the value of $\beta \Delta W$ transitions are less or more
likely. At very high temperature ($T \to \infty$) there is no bias; at
very low temperature ($T \to 0$) there is a strong bias. There is a
threshold value $T_c$ below which the color diffusion mode is unstable.
The order parameter is the difference in concentration of red and yellow
particles.

 Similar fluctuating bias fields are present in the {\em negative
viscosity}  model  of  Rothman [1989].  The fluctuating bias field is
the gradient of  the local flow field and the flux is the momentum flux.
This model leads to spontaneous ordering in the velocity field with
regions of high vorticity.  The order parameter is the transverse
momentum density or vorticity.

A further simplification of  the  Rothman-Keller model is the {\em
negative diffusion} model \cite{Alex}. It is defined on a square
lattice. Total mass per node is conserved, total momentum is not. The
model has only a single slow mode. The bias factor  $\exp[ \beta \vec
J(s^\prime) \cdot \vec G( {\bf s}_{n.n.})]$ contains the gradient of the
microscopic density $\rho(s(\vec r))$ and the particle current,
\begin{eqnarray} \label{b3}
  \vec G( {\bf s}_{n.n.}) & = & {\sum_m} \vec c_m \rho(s(\vec r
  +\vec c_m)) \nonumber \\
 \vec J(s^\prime)  & = & \sum_i \vec c_i s^\prime_i(\vec r)
\end{eqnarray}
The collection of models described above violate the condition of 
detailed balance.

We need to caution the reader here, because it is not at all clear
whether the forces driving the kinetics of phase separation  in
diffusive systems or the kinetics of ordering of  the flow field in
regions  of high vorticity, can be sensibly modeled by sending fluxes
upstream against the prevailing gradients (negative transport
coefficients). According to irreversible thermodynamics fluxes are
downstream, transport coefficients are positive, as is the irreversible
entropy production. These questions need further  clarification.

Here we will only illustrate some of the dramatic effects, such as
pattern formation, that results from negative transport coefficients
induced by fluctuating bias fields. 

As an illustration we discuss the negative diffusion model and apply the
linear stability analysis of section 2. For the special case of the {\em
half-filled} lattice ($\rho =2$) the eigenvalue equation (\ref{a3}) can
be solved analytically to yield the dispersion relation,
\begin{equation}\label{b4}
  z_D({\vec k})=\log[{\textstyle\frac{1}{2}}(\cos k_x+\cos k_y)
  + 4\omega (\sin^2 k_x+\sin^2 k_y)]
\end{equation}
for the diffusion mode, where $\omega$ is a typical matrix element of
the nonlocal collision operator, defined by Alexander et al.\ [1992]. 
We first apply the stability criterion of section 2.

\begin{figure}[t]
\centerline{
 \psfig{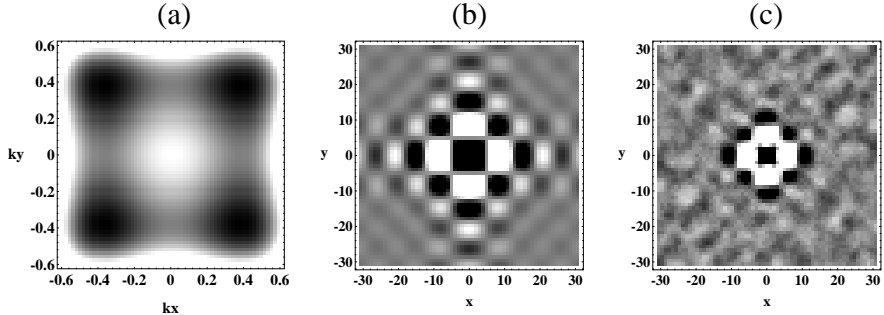}
}
\caption{Negative diffusion model at density $\rho=2$ and temperature 
$T=3.5$. No external driving field.
(a) Unstable (positive) parts of $z_D(\vec k)$ around origin of Brillouin 
zone, denoted by shades of gray.  Stable (negative) values are left blank. 
The spectrum has the symmetry of the square lattice.
(b) Density-density correlation function $G(\vec r,t)$ at $t=100$, obtained 
from the eigenvalues in (a) using Cahn-Hilliard theory. Maximal positive 
values are denoted by black, maximal negative values by white, and intermediate 
values by shades of gray. 
(c) Simulated values of $G(\vec r,t)$ at $t=100$, obtained from one run
for a $1024 \times 1024$ system at $T=3.0$.}
\end{figure}

The eigenvalue $ z_D({\vec k})$ is plotted in Fig.~1, for values of the 
control parameter above ($T=4.0$) and below ($T=3.5$) threshold. In Fig.~3a 
the unstable regions of $z_D({\vec k})$ in the $(k_x,k_y)$-plane are
indicated in shades of gray. The maxima are located in the
$45^o$-direction, a typical distance $k_m$ away from the origin. As $T$
is further decreased below threshold, the typical wavelength $\lambda_m
= 2 \pi/k_m$ of the most unstable excitation decreases. The threshold
value follows directly from the small-$k$ expansion of (\ref{b4}),
yielding  $z_D(\vec k)=-({\textstyle\frac{1}{4}}-4\omega) k^2 \equiv -Dk^2$.
For $\omega > \omega_c ={ 1/ 16}$ (which corresponds to $T <  T_c \simeq3.8$ 
according to Bussemaker and Ernst [1993c]) the diffusion coefficient takes 
a negative value, and the
system becomes unstable for long wavelength fluctuations. The mean
field value found for the threshold $(T_c \simeq 3.8)$ is about 20\% 
higher than the threshold measured in the computer simulations 
of Alexander et al.\ [1992]. 

According to the Cahn-Hilliard theory we can use (\ref{a7}) and
(\ref{a8}) to calculate the
initial structure and patterns in the density-density correlation
function $G(\vec r,t)$. This yields the checkerboard pattern of Fig.~3b
with a typical lattice distance $\lambda_m$. We have also carried out
computer simulations of the structure function and its inverse  $G(\vec
r,t)$ for a $1024 \times 1024$ system prepared in a random initial state 
at $t=0$. Quantitative comparison with the Cahn-Hilliard prediction of 
$G(\vec r,t)$ is complicated by the 20\% discrepancy in the threshold 
temperature, mentioned in the preceding paragraph.
Fig.~3c shows $G(\vec r,t)$ at $t=100$, obtained from a simulation at
$T=3.0$, a value which is just below the {\em simulation} value of the
threshold temperature $T_c=$3.05-3.10 reported by Alexander et al.\ [1992].
One sees that the structure of $G(\vec r,t)$ in the actual simulations is
qualitatively the same as in the Cahn-Hillard theory, although the typical
length scale is about 30\% smaller.

\section{Striped phases}

\subsection{Driven diffusive systems}

If the spatial symmetry of the underlying lattice is broken by an
external driving field or by a spatially uniform flow, the $\vec k$-regions 
of the most unstable excitations become strongly anisotropic,
and different types of striped phases may appear.

The occurrence of striped phases in driven diffusive systems is well
known \cite{Zia}. They have also been observed in computer simulations
of the negative diffusion
model, where an external field $\vec F$ is added, that drives a particle
current \cite{Alex}. The driving force in lattice gases can be
implemented by replacing the bias factor in section 3.3 by
$\exp[\,( \beta  \vec G( {\bf s}_{n.n.}) + \vec F) \cdot \vec
J(s^\prime)\,]$. 

In our analysis we restrict ourselves to the linear response regime,
where the particle current satisfies the constitutive relation,
\begin{equation} \label{c1}
  \vec J =  \mu \vec F + D \nabla \rho
\end{equation}
with $\mu$ the mobility and $D$ the diffusion coefficient. Both
transport coefficients are related by the Einstein relation $D=T \mu$.

For sufficiently high temperature  $T$ there exists a stable spatially
uniform state. Its distribution function $f^o_i(\vec F)$ can be obtained
by numerically solving (\ref{a1}). The velocity distribution is
anisotropic due to the presence of the field. As $T$ is decreased the
density fluctuations of long wavelength become again unstable. The
stability analysis of section 2 is performed by linearizing (\ref{a1})
around  $f^o_i(\vec F)$, and determining when and where the eigenvalue $
z_D({\vec k})$ becomes negative. As the spatial symmetry is broken the
threshold of stability $T_c(\vec F,\hat k)$ depends not only on the
field $\vec F$ but also on the direction of $\vec k$.

\begin{figure}
\centerline{
 \psfig{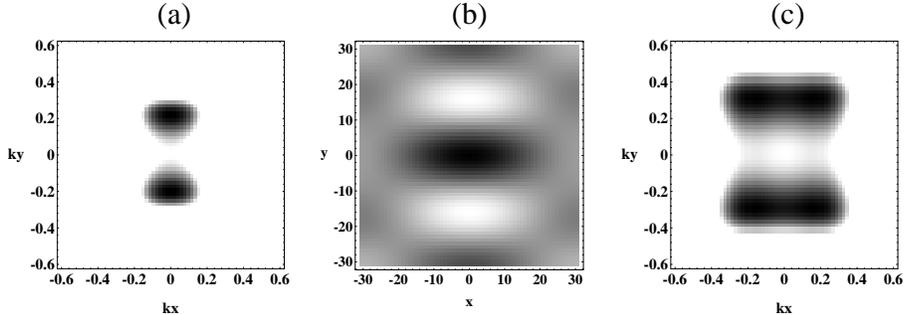}
}
\caption{Negative diffusion model at density $\rho=2$,
in the presence of an external driving field $\vec F=(0.5;0.0)$.
(a) Unstable (positive) parts of $z_D(\vec k)$ around origin of Brillouin 
zone, denoted by shades of gray, at $T=3.6$.  
Stable (negative) values are left blank. 
The symmetry of the square lattice is broken by the driving field.
(b) Density-density correlation function $G(\vec r,t)$ at $t=100$, obtained 
from the eigenvalues in (a) using Cahn-Hilliard theory.
(c) At $T=3.5$ the eigenvalue $z_D(\vec k)$ has become unstable in all 
$\vec k$-directions.}
\end{figure}

A systematic analysis of the $\vec k$-regions of instability as a function
of $T$ will be presented elsewhere \cite{BE-Alex}. Here we only illustrate
the occurrence of striped patterns for a typical case with $\vec F$ in the
$x$-direction.
At $T=3.6$ the eigenvalue $z_D({\vec k})$ only shows unstable modes if 
$\vec k$ is perpendicular to $\vec F$, with peaks at $|\vec k_m| \simeq 0.2$. 
Directions parallel to $\vec F$ are stable. See Fig.~4a.
By combining the numerical results for $z_D({\vec k})$ with the Cahn-Hilliard
theory in (\ref{a7}) and (\ref{a8}) one obtains the density-density
correlation function of Fig.~4b, indicating a striped pattern. 

Alexander et al.\ [1992] 
mention that ``there also appears to be a second transition at a
lower temperature to a phase with structures resembling those found for
${\vec F}=0$''.  Our analysis suggests that this transition is very
gradual: as the temperature decreases further below $T_c$, the region of
unstable ${\vec k}$-directions spreads out, until at some temperature all
directions have become unstable (Fig.~4c).  
Of course for very low temperatures the influence of ${\vec F}$ becomes 
negligible, since the term $\beta{\vec G}$ then dominates the bias factor.

\subsection{Unstable uniform flows}

In computer simulations on FCHC lattice gases H\'enon [1992]
has observed that an initially uniform flow is unstable. The flow orders
itself into parallel stripes  $\parallel \vec u_0$ with alternating flow
velocities parallel and anti-parallel to $\vec u_o$. The lattice gas
violates the Stueckelberg condition of semi-detailed balance. The above
observations  suggest that uniform flows tend to destabilize finite
speed equilibria.

To obtain some indications about possible instabilities in uniform flows
in lattice gases we simplify the nonlinear collision operator in (\ref{a1})
to a BGK-collision term \cite{BGK-LBE}, i.e.
\begin{equation} \label{c2}
  I_i = - \frac{1}{\tau} (f_i -f_i^\ell)
\end{equation}
where all non-vanishing eigenvalues $\omega_\lambda = 1/\tau$ are
related to  a single relaxation time $\tau$. The `local equilibrium'
distribution function $f_i^\ell$  in a $d$-dimensional lattice with
lattice distance $|\vec c_i| =v_o$ has been consistently chosen as,
\begin{eqnarray} \label{c3}
  f^\ell_i &=& f+ \rho \left(\frac{d}{b} \left(\frac{\vec c_i
  \cdot \vec u}{v^2_o} \right) + \frac{d(d+2)}{2b} \left( \frac{\vec c_i
  \cdot \vec u}{v^2_o} \right)^2 - 
  \frac{d}{2b} \left(\frac{u^2}{v_o^2} \right) \right) \nonumber \\
  f^\ell_0 &=& f_0 - \rho  \left(\frac{u^2}{v_o^2} \right) 
\end{eqnarray}
where $\rho = f_0 + bf$. This choice guarantees that the momentum flux
in local equilibrium is Galilei-invariant up to and including 
${\cal O}(u^2)$-terms, i.e. 
$ \Pi^\ell_{\alpha,\beta} = p \delta_{\alpha,\beta} + \rho u_\alpha
u_\beta $
where $\alpha$ and $\beta$ denote Cartesian components and $p=(
v^2_o/d)\,bf$ is the hydrostatic pressure, as in (\ref{b2}). This choice
still leaves $f_0$ as a free parameter to model the speed of sound
through $v^2_s = (v_o^2/d) [ 1 -df_0(\rho)/d \rho]$. Alternatively one
may add extra collision terms on the right hand side of (\ref{c2})
to model the mechanism of collisional transfer. 

In either case one can apply the method of section 2 to investigate the
stability of the uniform stationary state  $f^\ell_i(\vec u_o)$ with a
finite flow velocity $\vec u_o$. It turns out that the shear modes
remain stable, but the sound modes $(\sigma = \pm)$ may become unstable.
The sound damping constant is calculated as
\begin{equation} \label{c4}
  \Gamma_\sigma(\hat k) = \frac{1}{2v_s} (\tau - \textstyle{1\over 2})\,
  (v_s + \sigma u_{o \ell}) \left\{ {\displaystyle\frac{3 v_o^2}{d+2}} - 
  (v_s+\sigma u_{o \ell})^2 \right\} 
\end{equation}
where $u_{o \ell} =\hat k \cdot \vec u_o$. The shaded area in Fig.~5
gives the domain in the $(v_s, u_{o \ell})$-plane where both sound
modes are stable. The lines marked $(\sigma=+)$ and $(\sigma=-)$
denote the zero's of (\ref{c4}). 
The $\vec k$-direction parallel to $\vec u_o$ is the most unstable
\footnote{After completion of this research we received the preprint on
`Stability analysis of lattice Boltzmann methods' by J.D. Sterling and
S. Chen, in which stability diagrams similar to Fig.~5 have been
obtained by numerical solution of BGK-models.}.

If $u_{o \ell} =0$ we recover  the BGK-approximation to the results
(\ref{a6}), because all non-vanishing eigenvalues satisfy $\omega_\nu =
\omega_\zeta = 1/\tau$.  If $v_s > \, v_o [3/(d+2)]^{1/2}$ the sound
modes are unstable in all directions, and the BGK-lattice Boltzmann
equation constitutes a mathematical model for the unstable propagating
sound modes, discussed in section 3.2.

\begin{figure}[t]
\centerline{
 \psfig{figure=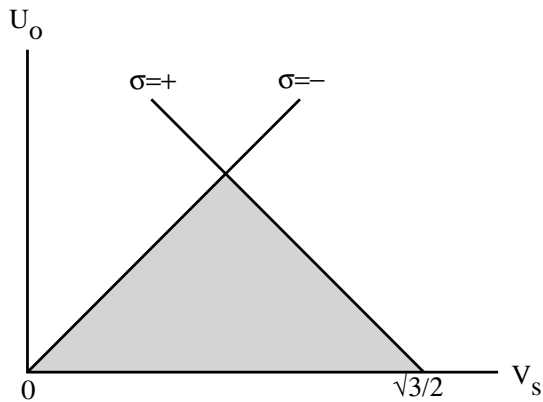,height=5.5cm}
}
\caption{Stability diagram for uniform flow in $(v_s,\,u_{o \ell})$-plane. 
 In the shaded region both sound modes are stable.
}
\end{figure}

Inspection of the diagram in Fig.~5 shows that it is easy to construct
BGK-models where sound waves with $\vec k \parallel \vec u_o$ are
unstable, and those with $\vec k \perp \vec u_o$ are stable.  In such
models the dynamic instability leads again to phase separation and
striped patterns, oriented perpendicular to the direction of the flow.
The order parameter is the sound mode, which is a linear combination of
momentum density parallel to $\vec u_o$ and mass density.

The diagram of Fig.~5 indicates how a stable basic equilibrium $(\vec u_o
=0)$ can become unstable when given a non-vanishing total momentum $\vec
P = N \vec u_o$, and how striped patterns can be formed with the sound
mode as order parameter. However, we  have not succeeded in constructing
a  BGK-model, that shows the H\'enon instability. Adding a flow velocity
$\vec u_o$ to the negative viscosity model of section 3.3 may lead to an
unstable shear mode at sufficiently high flow velocity, but generates
stripes perpendicular to the flow.\\[3mm]
{\em In summary}, we conclude that the stability analysis based on mean
field theory gives qualitative and quantitative information about
the initial patterns and accompanying length and time scales in phase
separation problems. The method presented here is applicable to both the
Boltzmann equation for lattice gas automata, and the mathematical
model Boltzmann equations, referred to as BGK-models.

\section*{Acknowledgements}

The authors acknowledge stimulating discussions with R. Brito, J.W.
Dufty, M. H\'enon and J. Somers. One of us (M.E.) thanks the Physics
Department of the University of Florida, where this research was
completed, for its hospitality in the summer of 1993, and acknowledges
support from a Nato Travel Grant for this visit. One of us (H.B.)
acknowledges support of the foundation `Fundamenteel Onderzoek der
Materie (FOM)',
which is financially supported by the Dutch National Science Foundation
(N.W.O.).

\end{document}